\documentclass{emulateapj}
\usepackage{apjfonts}
\begin{document}

\title{The Extended Star Formation History of the Andromeda Spheroid
at Twenty One Kiloparsecs on the Minor Axis\altaffilmark{1,2}}

\author{Thomas M. Brown\altaffilmark{3}, Ed Smith\altaffilmark{3}, 
Henry C. Ferguson\altaffilmark{3},
Puragra Guhathakurta\altaffilmark{4}, Jasonjot S. Kalirai\altaffilmark{4,5},
R. Michael Rich\altaffilmark{6}, 
Alvio Renzini\altaffilmark{7},
Allen V. Sweigart\altaffilmark{8},
David Reitzel\altaffilmark{6},
Karoline M. Gilbert\altaffilmark{6},
Marla Geha\altaffilmark{9}
}

\submitted{Accepted for publication in The Astrophysical Journal Letters}

\altaffiltext{1}{Based on observations made with the NASA/ESA Hubble
Space Telescope, obtained at STScI, and associated with proposal
10816.}

\altaffiltext{2}{Based on observations obtained at the W.M. Keck
Observatory, which is operated by the California Institute of
Technology, the University of California, and NASA, and made possible
by the generous financial support of the W.M. Keck Foundation.}

\altaffiltext{3}{Space Telescope Science Institute, 3700 San Martin Drive,
Baltimore, MD 21218;  tbrown@stsci.edu, edsmith@stsci.edu, ferguson@stsci.edu} 

\altaffiltext{4}{University of California Observatories / Lick Observatory, 
1156 High Street, University of California, Santa Cruz, CA 95064; 
raja@ucolick.org, jkalirai@ucolick.org, kgilbert@ucolick.org}

\altaffiltext{5}{Hubble Fellow}

\altaffiltext{6}{Department of Physics and Astronomy, 430 Portola Plaza,
Box 951547, University of California, Los Angeles, CA 90095;
rmr@astro.ucla.edu, reitzel@ucla.astro.edu}

\altaffiltext{7}{Osservatorio Astronomico, Vicolo Dell'Osservatorio 5, 
I-35122 Padova, Italy; arenzini@pd.astro.it}

\altaffiltext{8}{Code 667, NASA Goddard Space Flight Center, Greenbelt, MD
20771; allen.v.sweigart@nasa.gov}

\altaffiltext{9}{
NRC Herzberg Institute of Astrophysics,
5071 West Saanich Road, Victoria BC V9E 2E7,
Canada; Plaskett Fellow; marla.geha@nrc-cnrc.gc.ca}

\begin{abstract}

Using the {\it HST} ACS, we have obtained deep optical images of a
southeast minor-axis field in the Andromeda Galaxy, 21~kpc from the
nucleus.  In both star counts and metallicity, this field represents a
transition zone between the metal-rich, highly-disturbed inner
spheroid that dominates within 15~kpc and the metal-poor, diffuse
population that dominates beyond 30~kpc.  The color-magnitude diagram
reaches well below the oldest main-sequence turnoff in the population,
allowing a reconstruction of the star formation history in this field.
Compared to the spheroid population at 11~kpc, the population at
21~kpc is $\sim$1.3~Gyr older and $\sim$0.2~dex more metal-poor, on
average.  However, like the population at 11~kpc, the population at
21~kpc exhibits an extended star formation history; one third of the
stars are younger than 10~Gyr, although only a few percent are younger
than 8~Gyr.  The relatively wide range of metallicity and age is
inconsistent with a single, rapid star-formation episode, and instead
suggests that the spheroid even at 21 kpc is dominated by the debris
of earlier merging events likely occurring more than 8 Gyr ago.

\end{abstract}


\section{Introduction}

According to hierarchical models of galaxy formation, spheroids form
in a repetitive process during the merger of galaxies and
protogalaxies, while disks form by the slow accretion of gas between
these merging events (e.g., White \& Frenk 1991).  Hierarchical models
based on cold dark matter have shown great success in reproducing the
observable universe on scales larger than a Mpc, but these models can
lead to a ``missing satellite problem,'' with many more dwarf galaxies
predicted than are actually seen around the Milky Way (Moore et al.\
1999).  In answer to this problem, Bullock, Kravtsov, \& Weinberg
(2000) suggested that after the reionization of the universe,
photoionization suppressed gas accretion in subhalos, keeping most of
them dark-matter dominated, while a large fraction of the subhalos
that became dwarf galaxies were tidally disrupted into the halos of
their parent galaxies.  Grebel \& Gallagher (2004) countered that the
presence of ancient stars in all dwarf galaxies, along with their wide
variety of star formation histories, is evidence against a dominant
suppression from reionization.  In any case, recent semi-analytical
hierarchical models of spheroid formation have had some success in
modeling the spheroids and satellites of large galaxies (e.g., Bullock
\& Johnston 2005).

The Andromeda Galaxy (NGC~224, M31) offers an ideal laboratory for
testing these models.  Because M31 is inclined nearly edge-on
($12^{\rm o}$; de Vaucouleurs 1958) at a distance of 770 kpc (Freedman
\& Madore 1990), wide surveys can map its morphology, metallicity, and
kinematics (e.g., Ferguson et al.\ 2002; Kalirai et al.\ 2006), while
deep color-magnitude diagrams (CMDs) can reveal its star-formation history
(e.g., Brown et al.\ 2006; Olsen et al.\ 2006).  Here we present a new
deep CMD of a minor-axis field located 21~kpc from the M31 nucleus.
Within 15~kpc is the
metal-rich, highly-disturbed spheroid that has been the subject of
numerous ``halo'' studies since the work of Mould \& Kristian (1986).
Beyond 30~kpc, two independent groups have recently discovered the
spheroid more closely resembles a textbook spiral galaxy halo (Irwin
et al.\ 2005; Guhathakurta et al.\ 2005), with lower metallicities and
a $r^{-2}$ power-law profile.  Our new field falls in the transition
zone between these two distinct regions: well beyond the tidal
features present in the inner spheroid (Ferguson et al.\ 2002), but
within that part of the spheroid exhibiting relatively high
metallicities (Kalirai et al.\ 2006) and a de Vaucouleurs
surface-brightness profile (Durrell et al.\ 2004).

\section{Observations and Data Reduction}

Using the Advanced Camera for Surveys (ACS)
on the {\it Hubble Space Telescope (HST)},  
we obtained deep optical images of a
minor-axis field 1.5$^{\rm o}$ (21 kpc) from the M31 nucleus, at
$\alpha_{2000} = 00^h49^m05^s$, $\delta_{2000} = 40^{\rm
o}17^{\prime}37^{\prime\prime}$.  The field falls outside of the
disturbances seen in the Ferguson et al.\ (2002) star count map,
including those revealed in the Sobel-filtered version by Fardal et
al.\ (2006).  The field lies a few arcmin beyond the ``clean halo''
fields of Ferguson et al.\ (2005), also observed with ACS but with
less depth.  The 21~kpc field is part of a new ACS program 
sampling the outer spheroid of M31; the remaining fields,
falling $\sim$35~kpc from the nucleus, will be discussed in a later
paper.

From 9--28 Aug 2006, we obtained 8 hours of images in the F606W filter
(broad $V$) and 13 hours in the F814W filter ($I$) on the Wide Field Camera, 
with every
exposure dithered to enable hot pixel removal, optimal point spread
function (PSF) sampling, smoothing of spatial variation in detector
response, and filling in the detector gap.  Because our reduction
process is the same used by Brown et al.\ (2006), we will only briefly
summarize it here.  The images were registered, rectified, rescaled to
0.03$^{\prime\prime}$ pixel$^{-1}$, and coadded using the DRIZZLE
package (Fruchter \& Hook 2002), with rejection of cosmic rays and hot
pixels.  PSF-fitting photometry, using the DAOPHOT-II software of
Stetson (1987), was corrected to agree with aperture photometry of
isolated stars, with the zeropoints calibrated at the 1\% level.  The
final catalog contains $\approx$12,500 stars (Figure 1).  Our
photometry is in the STMAG system: $m= -2.5 \times $~log$_{10}
f_\lambda -21.1$.  For those more familiar with the ABMAG system,
ABMAG~=~STMAG~$-0.169$ for $m_{F606W}$, and ABMAG~=~STMAG~$-0.840$ mag
for $m_{F814W}$.  We performed extensive artificial star tests to
characterize the photometric errors and completeness in the catalog.
To avoid affecting the properties we were trying to measure, we added
only 500 artificial stars per pass, but by using thousands of passes,
the tests contain over 4 million artificial stars.

Spectroscopy of red giant branch (RGB) stars in our field provides
kinematic context for the population.  The velocities (Figure 1) were
measured as part of an ongoing Keck/DEIMOS survey of RGB stars in the
M31 spheroid (Gilbert et al., in prep.), sampling various radii to map
the surface brightness profile, kinematic structure, and metallicity
distribution.  For details on target selection, data reduction,
velocity fitting, and separation of foreground dwarfs from M31 giants,
see Gilbert et al.\ (2006).  Compared to the velocities in Brown
et al.\ (2006), Figure 1$a$ includes $\sim$50\% more stars that were
recovered by reprocessing the data from this field.

The CMD of our 21~kpc field is shallower and far less crowded than the
CMDs of the inner spheroid, tidal stream, and outer disk obtained by
Brown et al.\ (2006).  Ideally, in our current program we would obtain
the same depth and star counts, but the scarcity of stars in these
fields forced us to investigate the trade-off between depth and star
counts that could be achieved in a program of reasonable size.
Through simulations, we found that we could determine the predominant
star formation history even with one-tenth the number of stars and 0.2 mag less
depth.  The penalty is some loss of sensitivity to minority population
components (e.g., few percent bursts), which we demonstrate below.
Note that the loss of depth in our current observations is somewhat
mitigated by the smaller crowding errors.

Another distinction between the fields in the current program and
those in our earlier programs is the charge transfer inefficiency
(CTI) of the ACS CCD, due to radiation damage.  Brown et al.\ (2006)
could find no evidence of CTI in the images of the inner spheroid,
disk, and stream, presumably because the CTI was mitigated by the
bright sky and crowded fields filling the CCD traps.  In the current
images of the outer spheroid, the CTI appears as obvious streaks
trailing the stars.  The appearance of CTI can be attributed to the
fainter sky in August M31 observations (due to the larger Sun angle),
the far lower crowding, and the CCD being 2 years older.  We thus
apply a CTI correction to our photometry, using the algorithm of Riess
\& Mack (2005).  The correction makes stars at the turnoff ($m_{F814W}
\approx 29$~mag) 0.04~mag brighter, on average; at $m_{F814W} \approx
28$~mag it is 0.02~mag, while at $m_{F814W} \approx 30$~mag it is
nearly 0.08~mag.  Note that the CTI correction is not applied to the
artificial star tests, because artificial stars are not clocked across
the detector.  Because the correction makes the faint stars brighter,
it has a small but noticeable effect on the star formation history
fits (\S3); if we neglected the correction, the 21~kpc population would
appear to be 180 Myr older and 0.15 dex more metal-poor.

\section{Analysis}

In Figure 1 we compare the spheroid populations at 21~kpc and 11~kpc.
Both fields exhibit a broad velocity distribution around the M31
systemic velocity; neither field shows clear evidence of a dominant
kinematically-cold component, as might be expected if there were a
majority contribution from a single stream or from the disk.  It is
difficult to make a fair qualitative comparison of the populations in
these fields if
their CMDs are shown at their full depth.  Thus, in Figure
1$e$, we show a simulation of how the 11~kpc population would appear
if observed under the same conditions as the 21~kpc field
(crowding errors, exposure time, and star counts).  The simulation has
two components: for faint stars ($m_{F814W} \geq 26.5$~mag for $-0.9
\leq m_{F606W}-m_{F814W} \leq -0.1$~mag, and all stars at $m_{F814W}
\geq 28.0$~mag) the population is a realization of the best-fit
model for the 11~kpc field (Brown et al.\ 2006), but scattered
using the completeness and photometric errors derived for the 21~kpc field
(based on the artificial star tests); for bright
stars, the population is a random draw (without repeats) on
PSF-fitting photometry for the 11~kpc field, using a subset of the
images approximating the exposure time in the 21~kpc field.  
Two components properly account for the significant
differences in crowding and completeness for the faint stars
(the lower RGB, subgiant branch [SGB], and main sequence, which are
well-matched by the models), and accurately reproduce the bright stars
(the HB, asymptotic giant branch, and upper RGB, where the models
are not as accurate).
  
From the comparison of the 11~kpc and 21~kpc fields, it is clear that
the 21~kpc population is somewhat more metal-poor and older.  The
median RGB color is 0.02~mag bluer (at $m_{F814W} = 27$~mag) and the
RGB bump (immediately below the HB) is 0.2~mag brighter, both of which
indicate lower metallicities.  Despite these lower metallicities, the
SGB and main sequence turnoff are 0.1~mag fainter, indicating older ages.  The
HB stars fall mostly in the red clump, but the fraction of blue HB stars
is twice as large, implying both older ages and lower
metallicities.  The red clump in the 21~kpc field does not have the
obvious extension to brighter luminosities seen in the red clump of
the 11~kpc field, implying the metal-rich stars in the 21~kpc field do
not extend to the young ages seen in the 11~kpc field.  The luminosity
at the base of the red clump suggests that both fields are at
approximately the same distance, and we will assume that in the fits
below; if we assumed the distance to the 21~kpc population was larger
due to flattening of the spheroid, our fits for the 21~kpc field would
shift younger.  The RGB in the 21~kpc population is intrinsically 7\%
broader (after accounting for measurement errors); this is primarily
due to a larger fraction of metal-poor stars, but also due to the
shifting of metal-rich stars to older ages.  Still, the 11~kpc and
21~kpc populations are not completely distinct.  Both fields exhibit a
broad RGB, indicating a wide range of metallicities.  When compared to
the 47~Tuc ridge line ({\it curve}; Brown et al.\ 2005), both
fields exhibit a brighter SGB and turnoff, implying intermediate-age
stars are present.

To produce a quantitative fit to the star formation history in the
21~kpc field, we used the methodology of Brown et al.\ (2006) and the
Starfish code of Harris \& Zaritsky (2001).  The isochrones come from
VandenBerg, Bergbusch, \& Dowler (2006), but were transformed to the
ACS bandpasses (Brown et al.\ 2005); the isochrones do not
include core He diffusion, which, if present, would shift our ages $\sim$10\%
younger. We restricted the fits to the lower RGB, SGB, and
upper main sequence, where the random and systematic errors are
minimized.  Specifically, the fit region was bounded by $26.5 \leq
m_{F814W} \leq 30.0$~mag and $-0.9 \leq m_{F606W}-m_{F814W} \leq
-0.1$~mag, but a small $0.1 \times 0.2$~mag section was
masked where it overlaps with the blue HB.  The region and mask were
the same used in our earlier fits to the disk and stream (the
fit region for the inner spheroid extended 0.5~mag fainter).

The quantitative fits agree well with the
inspection of the CMDs.  Figures 1$f$ and 1$g$ show the best-fit star
formation histories for the populations at 11~kpc (Brown et al.\ 2006) and
21~kpc.  Although the 21~kpc population spans a wide range
of age and metallicity, it does not extend as young as the 11~kpc
population, and there is more weight in the old metal-poor population.
A third of the stars are younger than 10~Gyr;
fits that omit such stars are excluded at the 8$\sigma$ level, while
fits that omit stars younger than 6~Gyr are excluded at the 3$\sigma$
level.  Fits that omit stars more metal-rich than [Fe/H]$=-0.5$
are excluded at the 3$\sigma$ level,
but the omission of supersolar stars reduces the fit quality by less
than 1$\sigma$ (the fitting algorithm 
can compensate with more weight at or just
below solar metallicity).  On average, 
the 21~kpc population is 0.2 dex more metal poor than the
11~kpc population (consistent with Kalirai et al.\ 2006) and
1.3~Gyr older.  In
Figure 1$h$, we show the fit to the CMD shown in Figure 1$e$ (which
simulates 11~kpc population as it would appear if observed in the
conditions of the 21~kpc field).  Although the CMD in Figure 1$e$ is
significantly degraded (compared to that in Figure 1$c$), it is clear
that the predominant star formation history can still be recovered
(i.e., the star formation history in  
Figure 1$h$ looks very similar to that in Figure 1$f$).

\section{Summary and Discussion}

The M31 spheroid exhibits an extended star formation history at 11~kpc
and 21~kpc on the minor axis.  Compared to the 11~kpc field, the
21~kpc field is 0.2~dex more metal-poor and 1.3~Gyr older, on average,
but the most striking difference in these populations is the presence
of stars younger than 8~Gyr.  At 11~kpc, 26\% of the stars are younger
than 8~Gyr, but at 21~kpc, $\lesssim$5\% of the stars are (Figure 1$f$
and 1$g$).

Brown et al.\ (2006) found that the intermediate-age population at
11~kpc could not be explained by a large contribution from stars
currently residing in the disk or a chance intersection with the orbit
of the tidal stream.  Our current study yields an additional
constraint: the 11~kpc spheroid population does not look like a linear
combination of the stream and 21~kpc spheroid populations.  Although
the best fit to the stream population implies it is $\sim$1~Gyr
younger than the 11~kpc spheroid population (on average), only 20\% of
the stream population is younger than 8~Gyr; most of the stream stars
are 8--10~Gyr old (see Brown et al.\ 2006, Figure 13).  The
similarities between the stream and 11~kpc field imply pollution of
the inner spheroid by stream debris, consistent with current models
and observations of the stream and associated features
(e.g., Fardal et al.\ 2006; Gilbert et al., in prep.), 
but the larger fraction of stars younger than 8~Gyr
in the 11~kpc field may indicate that the inner spheroid is also
polluted with stars disrupted from the disk.  It would be
interesting to model the stream with a live disk that accounts for
both dynamical friction in the stream and disruption of disk stars.

Although we cannot tie any particular sub-population to any particular
cannibalized galaxy, the wide range of ages and metallicities in the
spheroid offer strong support to the hierarchical model, with the
spheroid comprised of dispersed satellite galaxies and possibly debris
from impacts through the disk.  However, 
our results are not necessarily consistent with all predictions
from hierarchical semi-analytical models.
Kalirai et al.\ (2006) have shown that there is a metallicity gradient
in the spheroid, such that it becomes more metal-poor at increasing
radii.  With only two points, one cannot draw a definite trend, but
our data are consistent with such a metallicity gradient, and also
indicate that the spheroid becomes older at increasing radii.  Hierarchical
models predict that the spheroid forms inside-out (e.g., Bullock \& 
Johnston 2005), so one might expect younger ages at larger radii.
The caveat is that 
the age distribution in a disrupted satellite might not be
strongly tied to the time of its disruption; the star formation
history of the tidal stream shows no indication that its progenitor
was disrupted within the last Gyr (Brown et al.\ 2006), despite n-body
simulations which point to this timescale (e.g., Fardal et al.\ 2006;
Font et al.\ 2006).  It will be interesting to explore the age
distribution beyond 30~kpc, to see if the trend for increasing age at
increasing radii continues.

\acknowledgements

Support for proposal 10816 is provided by NASA through a grant from
STScI, which is operated by AURA, Inc., under NASA contract NAS
5-26555.  We acknowledge support from NSF grants
AST-0307966/AST-0507483 (PG) and AST-0307931 (RMR), NASA/STScI grants
GO-10265/GO-10816 (PG, RMR), and NASA Hubble Fellowship grant
HF-01185.01-A (JSK).  We are grateful to P.\ Stetson for his DAOPHOT
code, to J.\ Harris for his Starfish code, and to K.\ Johnston, A.\
Font, and M.\ Fardal for interesting discussions.

\clearpage

\begin{figure}[ht]
\epsscale{1.2}
\plotone{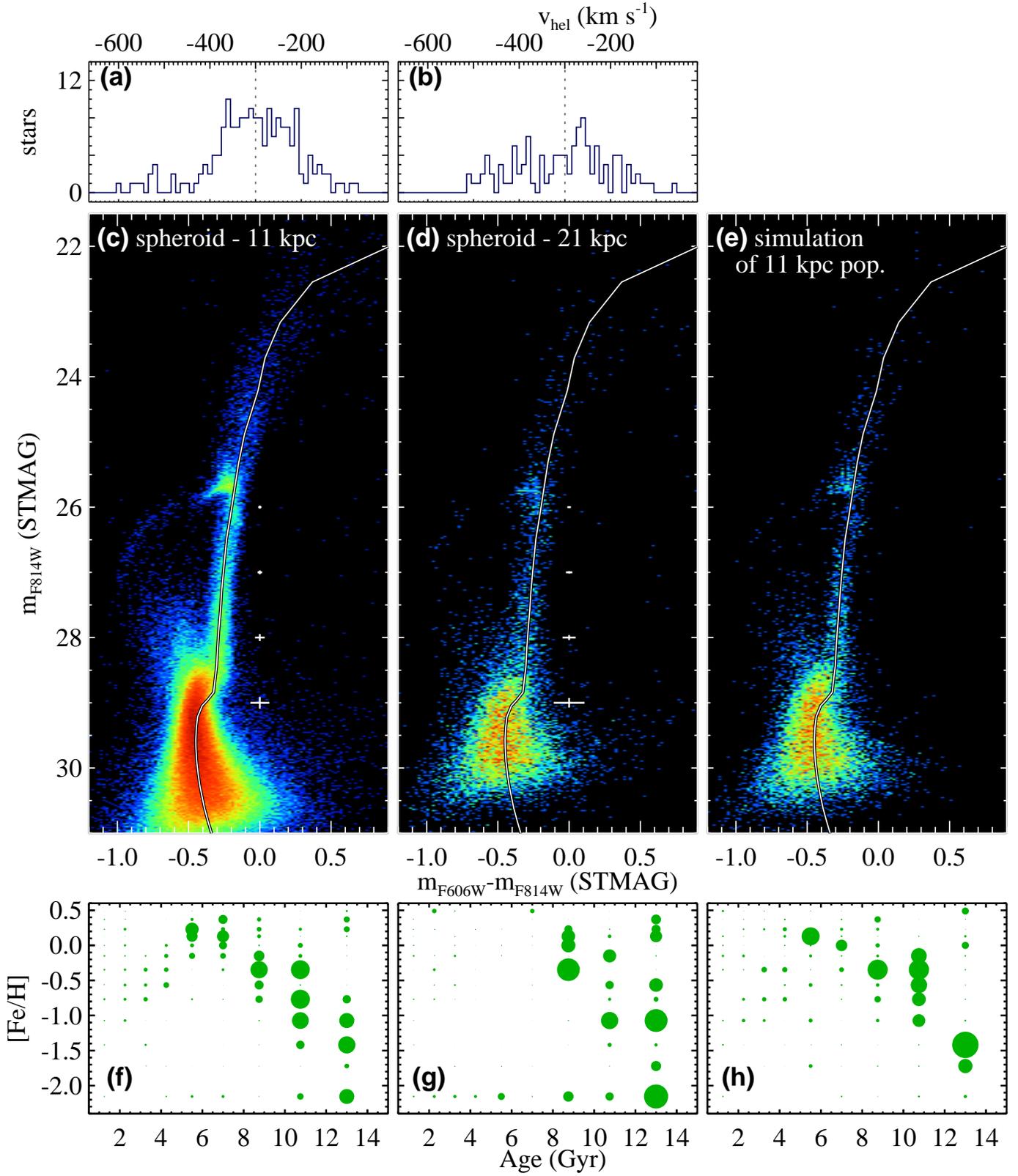} 
\epsscale{1.0}
\caption{{\it (a)} Velocities for RGB stars within 9$^\prime$ of our
11~kpc spheroid field, showing a broad distribution near the M31
systemic velocity ({\it dashed line}).  {\it (b)} Velocities for RGB
stars within 14$^\prime$ of our 21~kpc spheroid field, also showing a
broad distribution near the M31 systemic velocity.  {\it (c)} The CMD
of the spheroid population at 11~kpc.  The ridge line of NGC~104 ({\it
white curve}; Brown et al.\ 2005) is shown for comparison.  {\it (d)}
The same, but for the spheroid population at 21~kpc.  {\it (e)} A
simulation showing how the 21~kpc CMD (panel [$d$]) would appear if
its population were identical to that at 11~kpc.  {\it (f,g,h)} The
best-fit star formation histories for the CMDs shown in panels {\it
(c,d,e)}.  The area of each filled circle is proportional to the
number of stars at that age and metallicity.  The 21~kpc field hosts a
broad range of age and metallicity (panel [$g$]), but the population
is older and more metal-poor than that in the 11~kpc field (panel
[$f$]).}
\end{figure}

\end{document}